\documentclass{mem}
\usepackage{natbib}\usepackage{txfonts}\usepackage{balance}
\usepackage{graphicx}
\usepackage[a4paper]{hyperref}
\idline{75}{282}
\begin{document}
\def\teff{$T\rm_{eff }$}
\def\kms{$\mathrm {km s}^{-1}$}

\title{The Carina dSph galaxy: where is the edge?}

   \subtitle{}

\author{
G. Bono\inst{1}, M. Monelli\inst{1,2}, A.R. Walker \inst{3}, 
A. Munteanu\inst{4}, R. Buonanno\inst{1,5}, F. Caputo\inst{1}, 
V. Castellani\inst{1}, C.E. Corsi\inst{1}, M. Dall'Ora\inst{6}, 
P. Francois\inst{7}, M. Nonino\inst{8}, L. Pulone\inst{1}, 
V. Ripepi\inst{6}, H.A. Smith\inst{9}, P.B. Stetson\inst{10}, 
F. Thevenin\inst{11}
}

\offprints{G. Bono}

\institute{
INAF - Osservatorio Astronomico di Roma, Via Frascati 33, 00040 Monte Porzio 
Catone, Roma, Italy \email{bono@mporzio.astro.it};
\and
Instituto de Astrof\'isica de Canarias, Calle Via Lactea, E-38205 La Laguna, Tenerife, Spain;
\and 
Cerro Tololo Inter-American Observatory, NOAO, Casilla 603, La Serena, Chile;
\and
Universitat Pompeu Fabra, Dr. Aiguader 80, 08003 Barcelona, Spain;
\and
Universit\'a degli studi di Roma Tor Vergata, Via della Ricerca Scientifica 1, 00133 Roma, Italy;
\and
INAF, Sezione di Capodimonte, Via Moiariello 16, I-80161 Napoli, Italy;
\and
GEPI, Observatoire de Paris-Meudon, 92125 Meudon Cedex, France;
\and
INAF - Osservatorio Astronomico di Trieste, Via G.B. Tiepolo 11,
40131 Trieste, Italy; 
\and
Dept. of Physics, Michigan State University, East Lansing, MI 48824, USA;
\and
Dominion Astrophysical Observatory, Herzberg Institute of Astrophysics, 
NRC, 5071 West Saanich Road, Victoria, BC V9E 2E7, Canada,
\and
Observatoire de la C\^ote d'Azur, BP 4229, 06304 Nice Cedex 4, France;
}

\authorrunning{Bono et al.}

\titlerunning{The Carina dSph galaxy: where is the edge?}

\abstract{
Recent cosmological N-body simulations suggest that current
empirical estimates of tidal radii in dSphs might be underestimated
by at least one order of magnitude. To constrain the plausibility
of this theoretical framework, we undertook a multiband ($U$,$B$,$V$,$I$) 
survey of the Carina dSph. Deep $B$,$V$ data of several fields located at 
radial distances from the Carina center ranging from 0.5 to 4.5 degrees 
show a sizable sample of faint blue objects with the same magnitudes 
and colors of old, Turn-Off stars detected across the center.\\  
We found that the ($U$-$V$,$B$-$I$) color-color plane is a robust 
diagnostic to split stars from background galaxies. Unfortunately, 
current $U$,$I$-band data are too shallow to firmly constrain the real 
extent of Carina.  

}
\maketitle{}


\section{Introduction}
The Carina dSph plays a fundamental role among the dwarf galaxies 
in the Local Group, because it is relatively close to the Galaxy and 
shows multiple star-formation episodes \citep{smecker94}. 
Detailed Color-Magnitude Diagrams (CMDs) show an old ($t\approx 11$ Gyr)
stellar component, including sub-giant and Horizontal Branch (HB) stars, 
an intermediate-age ($t\approx 5$ Gyr) component including Turn-Off 
and Red Clump stars, and a younger ($t\le 1$ Gyr) component of blue 
Main Sequence (MS) stars \citep{smecker96, cicci, io}.

Photometric surveys based on robust stellar tracers (RR Lyrae, Red Giants) 
indicate the existence of extra-tidal stars \citep{kuhn, majewski-extratidalI,
majewski-extratidalII}. However, this empirical evidence is 
hampered by small-number statistics. This is the reason why we decided 
to use MS stars bluer than $B-V$=0.4 to trace the radial extent of this 
galaxy. Recent predictions based on detailed N-body simulations suggest 
that empirical tidal radii of dSphs might be significantly larger than 
currently estimated. In particular, \citet{hayashi} found that 
the Carina tidal radius might be at least one order of magnitude larger 
than estimated by \citet{majewski-extratidalI}. This prediction is further 
supported by independent calculations by \citet{mayer} who found
that the current luminosity cut-off of Carina is too small when 
compared with the predicted massive dark halo of this dSph.

To assess the existence of extra-tidal stars around Carina, we undertook
a detailed study of the outer regions of this galaxy collecting deep, wide
field, and multiwavelength data. In the following we discuss some preliminary 
results based on a subset of these data. 

\section{Observations and preliminary results}

We observed the central region and eight different fields along the major 
and minor axes, with distances ranging from 0.5 up to 4.5 degrees from 
the Carina center. These regions have been observed in $B$ and $V$ bands 
with the MOSAICII camera (f.o.v. 36$^\prime\times$36$^\prime$) available 
at the 4m CTIO Blanco telescope. 

Time series data of the Carina center were collected in December 1999 and 
January 2000. The external fields was observed in different runs between 
October 2002 and January 2005.  Standard IRAF routines \citep{valdes} were 
adopted for basic reduction, and the photometric analysis was performed 
using DAOPHOT/ALLFRAME \citep{stetson-dao,stetson-allf}. The details of 
the reduction and calibration strategy will be discussed in a forthcoming 
paper (Monelli et al. 2005, in prep.).

   \begin{figure}
   \centering
   \includegraphics[width=5.5cm,height=7cm]{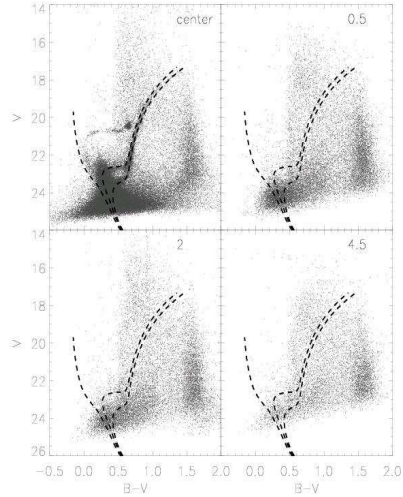}
    \vspace{0.6cm}
   \caption{\footnotesize {\bf Top Left} - CMD ($V$,$B-V$) of the 
Carina central region ($36\times36$ arcmin) which includes 
$\approx 90,000$ objects. The comparison with stellar isochrones 
(DM=20.24, E(B-V)=0.03) suggests  
that the three main star formation episodes occurred at $t\approx$11, 5, 
and 1 Gyr ago. {\bf Others} - The same isochrones have been overplotted 
on the CMDs of three external regions located at different distances 
ranging from 0.5 to 4.5 degrees from the center. Note that the spur of 
faint blue objects located at $23 \le V\le24.5$, $B-V\le 0.4$ is located 
in the same CMD region of the Carina old MS stars. The different CMDs 
include on average $\approx 20,000$ objects.  
   }
   \label{one}
   \end{figure}
%
%
Fig. \ref{one} shows the Carina $V$,$B$-$V$ CMDs of the central (top 
left panel) and of outer (other) regions. The CMDs of the external fields 
disclose a sizable sample of faint blue objects ($23 \le V\le24.5$, 
$B-V\le 0.4$). 
The comparison between observations and isochrones \citep{sait2004},  
at fixed chemical composition (DM=0.24, Z=0.0004), indicates that these 
objects are located in the same CMD region of old TO stars we have already 
detected in the Carina center. Note that this spur of faint blue objects 
is present in all the fields we observed, up to a distance of 4.5$^\circ$ 
from the center. 

In order to explain the
nature of these objects we are left with three working hypotheses: 

{\em i)} {\em Extra-tidal stars} - They could be either extra-tidal stars 
as originally suggested by \citet{kuhn} and by \citet{majewski-extratidalI} 
or belong to an extended halo surrounding Carina as suggested by recent 
N-body simulation \citep{hayashi, mayer, kaza}

{\em ii)} {\em Galactic field stars} - We performed several numerical 
simulations of Galactic models 
\citep{cast02} by adopting a broad range of input parameters, 
namely, the Initial Mass Function, the Star Formation Rate, and the 
thin/thick disk scale heights. Interestingly enough, we found that the 
contamination of field stars in the CMD region located between
$23 \le V\le 24.5$ and $0.0 \le B-V\le0.4$ is limited to small samples
of thin and thick disk white dwarfs. Fortunately, field halo stars in
the same magnitude range are systematically redder than the faint blue 
objects we have detected. 

%
   \begin{figure}
   \centering
   \includegraphics[width=7cm,height=8cm]{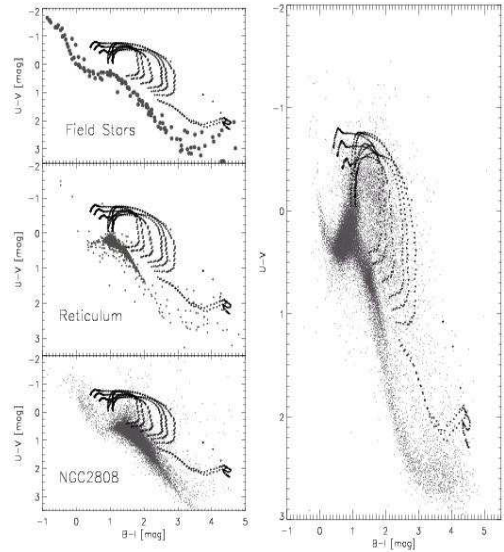}
   \caption{\footnotesize {\bf Left} - Color-color plane, $U$-$V$,$B$-$I$, 
showing the comparison between three stellar samples with metal abundance 
ranging from solar (top, [Fe/H]=0) down to [Fe/H]=-1.7 (bottom, NGC2808), 
and evolutionary sequences for galaxies with redshift $\le$2. 
{\bf Right} - Same as the right, but for the Carina central regions. 
   }
   \label{two}
   \end{figure}
%
{\em iii)} {\em Background galaxies} - To constrain on a quantitative basis the contamination by 
background galaxies, we devised a new diagnostic based on the ($U-V$,$B-I$) 
color-color plane. The left panels of Fig. \ref{two} shows the comparison
between different stellar templates, namely a sample of field stars, two
globular clusters (Reticulum [Large Magellanic Cloud], NGC2808 [Galaxy]) 
for which accurate multi-band data are available, and predicted colors 
for background galaxies provided by \citet{fioc}.
Data plotted in this figure display that TO stars ($U-B\approx0.5$,
$B-I\approx1$) are systematically redder than background galaxies
at redshift smaller than 2 \citep{fontana}.
Therefore, we collected $U$,$I$-band data with the MOSAICII 
camera available at the 4m CTIO telescope of the central field 
and of a field located 1 degree southern from the Carina center.

%
   \begin{figure}
   \centering
   \includegraphics[width=6cm,height=7cm]{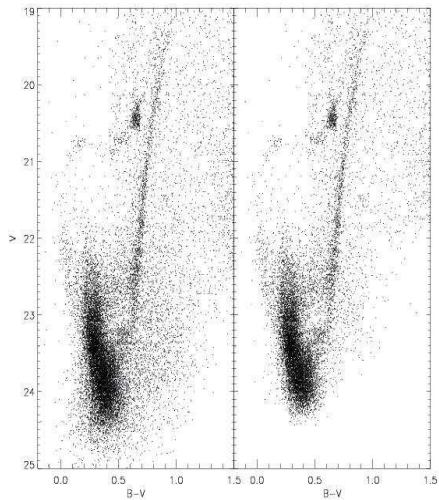}
   \caption{\footnotesize  The left and the right panel show the CMD of 
the Carina central regions before and after the cleaning from galaxy 
contamination. Approximately $\approx 6000$ galaxy candidates have been 
selected from the color-color plane and subtracted. The left panel shows 
the objects ($\approx$40000) which have been detected in all the four 
bands. The TO region of the old stellar component appears better defined 
in the right panel and the spread in color of MS stars is smaller.
}
   \label{three}
   \end{figure}
%
Data plotted in the right panel of Fig. \ref{two} show the color-color plane 
of the Carina central region. We selected $\approx$6000 objects in 
the region where background galaxies are expected. Fig. \ref{three} shows the 
Carina $V$,$B$-$V$ CMD before (left) and after (right) subtracting 
background galaxies. This diagnostic appears to work quite well, since down 
to $V\sim23.5$ the $\approx$70\% of blue objects appear to be real 
stars. Moreover, the TO region seems better defined, and the color 
spread of MS stars is smaller. The width in color of MS in dSphs 
has been considered as a robust evidence of a spread both in age 
and/or in chemical composition. Data plotted in Fig. \ref{three} indicate 
that this region is also contaminated by background galaxies. 
The removal of these objects will allow us to better constrain the 
TO of the different populations, and in turn to provide robust estimates 
of the different star formation episodes.

The limiting magnitude of $U$ and $I$-band data is too 
shallow to firmly establish the nature of the objects around the 
TO luminosity of old MS stars ($V\sim24.5$, $U\sim25$). Unfortunately, 
this problem is even more severe for the field located at 1 degree from 
the Carina center (bad weather conditions). However, the circumstantial 
evidence that the blue objects we detected are located in a region of 
the color-color plane typical of MS stars indicates that they might 
be truly Carina stars.


\section{Final remarks} 

The preliminary results presented in this investigation are part of a 
long-term photometric and spectroscopic project aimed at investigating 
the stellar content of the Carina dSphs (Monelli et al. 2005, in prep.). 
We have already collected photometric data over a large area around Carina 
and both low and intermediate-resolution spectra across the center. 
However, accurate estimates of a basic parameter such as the tidal radius 
still hinge on the robust multiband identification of the Carina faint 
stellar components.  In particular, it appears crucial to establish 
whether both old and intermediate-age, extra-tidal stars are present 
in Carina, and in turn whether they are distributed in an extended 
spherical halo or along tidal stream(s). 
These occurrences will supply robust empirical constraints on the 
physical assumptions currently adopted in numerical simulations of 
galaxy formation and evolution.

No doubt that a comprehensive photometric and spectroscopic 
investigation of the Carina stellar 
structure will be an important step forward in our knowledge of 
these elusive stellar systems and how they interact with the Galaxy.

\begin{acknowledgements}
This work was partially supported by MIUR/COFIN003. 
\end{acknowledgements}

\bibliographystyle{aa}

\end{document}